\documentclass[conference,a4paper]{IEEEtran}
\usepackage{cite}
\usepackage{flushend}
\usepackage[dvips]{graphicx}

\usepackage[cmex10]{amsmath}
\usepackage{gensymb,amsfonts,amssymb}
\usepackage{bm,array}
\usepackage{algorithmic}

\usepackage{array}
\usepackage[caption=false,font=footnotesize]{subfig}
\usepackage{fixltx2e,soul}

\usepackage{url}
\usepackage{xcolor}
\begin{document}
\title{Beamwidth Optimization in Millimeter Wave Small Cell Networks with Relay Nodes: A Swarm Intelligence Approach}


\author{\IEEEauthorblockN{Cristina Perfecto\IEEEauthorrefmark{1}, Javier Del Ser\IEEEauthorrefmark{1}\IEEEauthorrefmark{2}, Muhammad Ikram Ashraf\IEEEauthorrefmark{3}, Miren Nekane Bilbao\IEEEauthorrefmark{1}, Mehdi Bennis\IEEEauthorrefmark{3}}
\IEEEauthorblockA{\IEEEauthorrefmark{1}University of the Basque Country - UPV/EHU, Spain\\ Email: \{cristina.perfecto, javier.delser, nekane.bilbao\}@ehu.eus}
\IEEEauthorblockA{\IEEEauthorrefmark{2}TECNALIA RESEARCH \& INNOVATION, Spain\\ Email: javier.delser@tecnalia.com}
\IEEEauthorblockA{\IEEEauthorrefmark{3}Centre for Wireless Communications - CWC, University of Oulu, Finland \\Email: \{ikram, bennis\}@ee.oulu.fi}}
\maketitle

\begin{abstract}
Millimeter wave (mmWave) communications have been postulated as one of the most disruptive technologies for future 5G systems. Among mmWave bands the 60-GHz radio technology is specially suited for ultradense small cells and mobile data offloading scenarios. Many challenges remain to be addressed in mmWave communications but among them deafness, or  misalignment between transmitter and receivers beams, and interference management lie among the most prominent ones. In the recent years, scenarios considering negligible interference on mmWave resource allocation have been rather common in literature. To this end, interestingly, many open issues still need to be addressed such as the applicability of noise-limited regime for mmWave. Furthermore, in mmWave the beam-steering mechanism imposes a forced silence period, in the course of which no data can be conveyed, that should not be neglected in throughput/delay calculations. This paper introduces mmWave enabled Small Cell Networks (SCNs) with relaying capabilities where as a result of a coordinated meta-heuristically optimized beamwidth/alignment-delay approach overall system throughput is optimized. Simulations have been conveyed for three transmitter densities under TDMA and na\"{\i}ve `all-on' scheduling producing average per node throughput increments of up to 248\%. The paper further elaborates on the off-balancing impact of alignment delay and time-multiplexing strategies by illustrating how the foreseen transition that increa\-sing the number of transmitters produces in the regime of a fixed-node size SCN in downlink operation fades out by a poor choice in the scheduling strategy.
\end{abstract}
\IEEEpeerreviewmaketitle

\section{Introduction}

In recent years the 60-GHz band has garnered renewed interest among the research community due to its great potential to meet the increasing demand for affordable and ultra high-speed data relay connectivity. This spectrum band was initially considered exclusively for indoor use and wireless personal area network (WPAN) communications with coverage distances of only a few meters. The main cause for such a consideration being oxygen's molecular resonance causing signals to fade much more rapidly with distance as compared to the current UHF/microwave bands. However, first the allocation of 7GHz of spectrum for unlicensed operation between 57-64 GHz by the Federal Communication Commission (FCC), and second the FCC regulations revision of Part 15 rule allowing for higher transmission power for 60-GHz devices that operate outdoors \cite{FCCRule152013}, have made possible to balance the large attenuation factor by means of highly directive antenna gains in link budget calculations, which overcome the plight to produce large amounts of radio frequency power at mmWave with low-cost integrated circuits. This growing interest is further buttressed by the number of standardization activities --ECMA 387 (2008), IEEE 802.15.3c (2009), WiGig (2011), IEEE 802.11ad (2012)--, study groups around WPAN (last IEEE 802.11ay established in May 2015) or EU H2020 5G-PPP projects for mmWave fronthaul, backhaul and small cell access in ultra-dense deployments in future 5G heterogeneous cellular networks (e.g. MiWave\cite{EU-FP7-MiWaves}, mmMagic\cite{EU-5GPPP-mmMagic}).

\subsection{Related Work}\label{subsec:related}

In mmWave band the use of highly directional antennas drastically reduces interference but it also leads to ``deafness" if transmitter and target receiver antenna boresights are not correctly oriented towards each other. In mmWave SCNs small cell base stations (SCBSs) and user equipments (UEs), endowed with steerable antennas, will have to undertake a beam alignment period, also referred as \textit{beam-searching} or \textit{beam-training}, before communication. The beamforming itself can be achieved by adopting analog, digital or hybrid architectures. Until hybrid analog-digital beamforming (allowing for reduced complexity while still providing multiplexing gains) takes over in forthcoming standards like IEEE 802.11ay, analog beamforming architectures remain the choice for current mmWave WPAN standards such as IEEE 802.11ad \cite{ieee:802.11ad_part11} and IEEE 802.15.3c \cite{ieee:802.15.3.c_part15.3}. Similarly, albeit numerous alternatives that speed-up the beamforming protocol have been proposed in the recent literature, a simplified version of the 3-step beam codebook-based approach introduced by \cite{Wang2009} is employed due to its robustness, simplicity and compliance with ongoing standards. Through a sequence of pilot transmissions and a trial-and-error approach, the two-staged beam alignment process (encompassed by an initial coarse sector-level scanning and followed by a finer granularity beam-level search within the limits of elected sector) yields refined beams orientation. As all possible combinations need to be explored, search space scales and so does induced overhead with 1) device mobility --that would lead to sector hops-- and 2) with the product of sender-receiver beam resolution.

Additionally, numerical results on interference analysis in highly directional links for outdoor mmWave networks in \cite{Singh2011} lead to the idea of a generalized \emph{pseudowired} link abstraction governed by a noise-limited regime. By contrast, different mechanisms for interference mitigation in the context of mmWave are presented in \cite{Park2009,Cai2012}. More recently, \cite{Shokri-Ghadikolaei2015d} called into question the common belief that indoor mmWave WPANs were also mainly noise-limited. It was argued that, either resorting to mimicking MAC protocols from omnidirectional networks whose focus is set on interference management or, alternatively, essentially disregarding the effect of interference, accounted for current mmWave systems being significantly less bandwidth efficient than expected and producing data rates still one order of magnitude below the 100 Gbps that IEEE 802.11ay foresees for short range networks.

\subsection{Contributions}\label{subsec:contribution}

In this paper the above research line is addressed from a quantitative approach by studying how a fixed mmWave SCN with relaying capabilities behaves for different relay UE densities under optimized transmission and reception operating beamwidths. It is shown that a Swarm Intelligence algorithm, specifically Particle Swarm Optimization (PSO), converges to a solution that maximizes the average node throughput and evinces the existence of an interplay between the beam-training alignment overhead and the maximum achievable transmission rate \cite{Shokri-Ghadikolaei2015}. In particular, the PSO beamwidth optimization process illustrates that unless beamwidths for each transmitter-receiver pair are carefully chosen, situations in which a non-negligible multiuser interference (MUI) is present do happen even when very conservative resource allocation protocols, such as TDMA that activates only one link at a time, are adopted in the transmitters. Moreover, optimization results unveil that the burden imposed by scheduling and alignment delay into throughput calculations exceeds that of noise power or interference that seem no longer the limiting factors. Finally, for each transmitter density, average achievable throughputs and interference levels under a holistic `all on' scheduling operation are provided as a reference.

The rest of the paper is organized as follows: Section II delves into the system model and problem formulation. For the sake of completeness, Section III briefly oversees the fundamentals of our proposed PSO solver approach and simulation setup. Section IV presents and discusses simulation results. Finally, Section V concludes the paper by drawing conclusions and outlining some related lines of future research.

\begin{table}[t!]
	\renewcommand{\arraystretch}{1.2}
	\caption{Summary of Main Notations}\label{tab:Notation}
	\centering
	\begin{tabular}{|l|l|}
		\hline \textbf{Symbol}&\textbf{Definition}\\
		\hline SINR$_j$& Signal-to-interference-plus-noise-ratio in receiver j of link i\\	
		\hline$Z$ & Number of simultaneous transmitter-receiver pairs\\
		\hline$p_{i}$ & Transmission power of reference transmitter\\
		\hline$p_{z}$ & Transmission power of interfering transmitter\\
		\hline$g_{i,j}^c$ & Channel gain in link from transmitter i receiver j\\
		\hline$g_{i,j}^t$, $g_{i,j}^r$ & Antenna gain transmitter/receiver of link i\\
		\hline$N_0$ & Gaussian background noise power density\\
		\hline$\psi_i^t$, $\psi_i^r$ & Sector-level beamwidth transmitter/receiver of link i\\
		\hline$\varphi_i^t$, $\varphi_i^r$ & Beam-level beamwidth transmitter/receiver of link i\\
		\hline$\theta_{i,j}^t$, $\theta_{i,j}^r$ & Deviation angle relative to transmitter/receiver boresight\\
		\hline$\mathcal{B}_{mmW}$ & Bandwidth in mmWave band\\
		\hline$T$& Transmission Slot duration\\
		\hline$T_p$& Pilot transmission duration\\
		\hline$\tau_{i,j}$& Antenna alignment delay\\
		\hline
	\end{tabular}
\end{table}

\section{System Model and Problem Formulation}\label{sec:system-model}

Consider a downlink (DL) transmission in a mmWave SCN operating in Time Division Duplexing (TDD), where SCBS retrieves content from backhaul and delivers it to UEs through infrastructure-to-device (I2D) links. UEs are further separated into anchor UEs (AUEs) --an anchor node is considered as a serving UE-- and client UEs (CUEs); the former relay traffic of SCBS to CUEs inside the network by exploring device-to-device (D2D) links, whereas the latter may get those contents either directly from the SCBS or via an AUE. A co-channel deployment with bandwidth $\mathcal{B_\textup{mmW}}$ and uniform transmit power is assumed. Moreover, half-duplex UEs are considered in our model. It is also hypothesized that the network topology and channel conditions remain unchanged in the given time slot. Due to slow-varying channel state information (CSI) our model is suited for mmWave scenarios with low user mobility and line-of-sight (LOS) transmission.

Under the premise that SCBS acts as a coordinator and is aware of the network topology (node locations and appointed AUEs), SCBS could dictate the exact angle/steering for each transmitter-receiver pair and have them using the narrowest possible beams. However, applying this narrowest beam stra\-tegy would make a poor choice for most of the links: even in low mobility scenarios UEs are prone to small, yet significant, orientation fluctuations that could make pencil-like antenna beams off their target nodes and consequently need regular intervention from SCBS. Hence, SCBS would rather optimize beamwidths of each transmitter-receiver link in such a way that it jointly distributes the interplay between throughput, alignment delay resulting from legacy beam-training, the effect of interference treatment and MAC scheduling, so as to find a longer-term successful beam alignments.

\begin{figure}[h!]
	\captionsetup{farskip=2pt,captionskip=1pt}
	\centering
	{\includegraphics[width=.75\columnwidth]{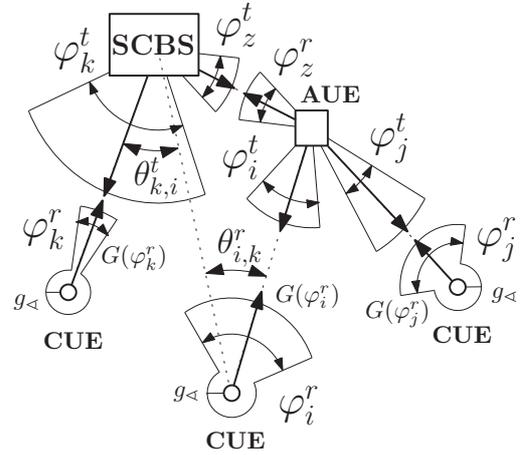}}
	\caption{MmWave SCN: Beam, angle and antenna gain details.}
	\label{fig:SCN_angles}
\end{figure}

An ideal sector antenna model that effectively captures key features of typical antenna array patterns \cite{Bai2015} is considered at both the SCBS and the UEs. This ideal sector antenna, as given by \eqref{eq:antenna} wherein $g_{i,j}^t$ and $g_{i,j}^r$ stand for transmitter and receiver directivity gains and $\theta_{i,j}$ for the angular deviation between receiver $j$ and transmitter $i$ relative to their respective boresight directions, entails a constant gain $G(\varphi_{i}^\wp)$, $\wp\in\{t,r\}$, for all emission angles within mainlobe, and a non-negligible sidelobe power $g_\sphericalangle$, with $0\leq g_\sphericalangle\ll 1$, outside:
\begin{equation}\label{eq:antenna}
g_{i,j}^\wp=\left\lbrace
\begin{array}{ll}
\frac{2\pi-(2\pi-\varphi_{i}^\wp)g_\sphericalangle}{\varphi_{i}^\wp}\text{,} & \text{if } |\theta_{i,j}^\wp|\leq \varphi_{i}^\wp/2,\\
\vspace*{-0.1cm}
g_\sphericalangle, &\text{otherwise.}\\
\end{array}	\right.\\
\end{equation}

According to Shannon theory, the maximum achievable data rate at receiver $j$ is determined by its measured signal to interference plus noise ratio (SINR), which hinges not only on the received signal power itself but also on interference and noise levels. The SINR for node $j$ is given by
\begin{equation}\label{eq:SINR_mm}
	SINR_{j}^{mmW}=\frac{p_ig_{i,j}^tg_{i,j}^cg_{i,j}^r}{\sum\limits_{\substack{z=1\\z\neq i}}^Z p_zg_{z,j}^tg_{z,j}^cg_{z,j}^r +N_0\mathcal{B_\textup{mmW}}},
\end{equation}
where $g_{i,j}^c$ is the channel gain, $p_i$ the transmission power of transmitter $i$, $Z$ the number of simultaneous transmitter-receiver pairs each interfering on link $i$ with power $p_z g_{z,j}^t g_{z,j}^c g_{z,j}^r$, and $N_0$ the Gaussian noise power density.

In regards to SINR calculations MUI will be computed under the ``physical model'' \cite{Gupta2000} assumption where, in spite of marginal contributions from each interfering node being individually too weak a packet loss will still be deemed feasible if the overall sum of interferences exceeds a given threshold. MAC scheduling in the SCBS and in each AUE will limit the maximum number of simultaneous active links and, in kind reduce interfering sources for each receiver. In such cases, for reference receiver SINR calculation purposes worst interference source link from each interferer transmitter is considered in simulations.

\subsection{Alignment Delay and Transmission Rate}\label{sec:system-model-Delay/transmission}

Despite the abundant literature on mmWave beam searching, to the best of our knowledge no previous research other than \cite{Shokri-Ghadikolaei2015} has hitherto incorporated the alignment delay into the effective transmission rate. Let $T_p$ denote the pilot transmission time, and let the tuples $\{\psi_i^t$,$\varphi_i^t\}$, $\{\psi_i^r$,$\varphi_i^r\}$ represent sector and beam level widths of link $i$ at transmitter and receiver ends. Without loss of generality, we assume that sector level alignment has been established prior to beam-level alignment phase. So, by applying a continuous approximation \cite{Shokri-Ghadikolaei2015}, the alignment time $\tau_{i,j}$ is conveyed as
\begin{equation}\label{eq:align_time}
\tau_{i,j}(\varphi_i^t,\varphi_i^r)=\frac{\psi_i^t\psi_i^r}{\varphi_i^r\varphi_i^t}{T_p}.
\end{equation}

Moreover, $\tau_{i,j}$ cannot exceed transmission slot's duration $T$, hence a lower bound on the beamwidths product can be settled as
\begin{equation}\label{eq:low_beamwidth_product}
{\varphi_i^t}{\varphi_i^r}\geq\frac{T_p}{T}{\psi_i^t}{\psi_i^r}.
\end{equation}

Likewise, inasmuch as beam level cannot exceed sector-level widths and antenna configuration limits apply, upper and lower bounds on the operating beamwidth for transmitter and receiver are $\varphi_{i_{min}}^t\leq \varphi_i^t\leq\psi_i^t$ and $\varphi_{i_{min}}^r\leq\varphi_i^r\leq\psi_i^r$ respectively. Once the beams have been aligned, the achievable data rate over the remaining $T-\tau_{i,j}$ seconds normalized by time slot duration T, is given by \cite{Shokri-Ghadikolaei2015}.
\begin{equation}\label{eq:Eff_Rate}
R_{(i,j)}^{mmW}=\left(1-\frac{\tau_i}{T}\right)\mathcal{B}_{mmW}\log_2\left(1+SINR_{j}^{mmW}\right).
\end{equation}

\subsection{Problem Formulation}\label{sec:system-model-ProblemFormulation} 

Let ${\mathcal{K}}=\{1,\ldots,K\}$, ${\mathcal{L}}=\{1,\ldots,L\}$, ${\mathcal{M}}=\{1,\ldots,M\}$, ${\mathcal{N}}=\{1,\ldots,N\}$, ${\mathcal{K}=\mathcal{M}\cup\mathcal{N}}$, ${\mathcal{N}\cap\mathcal{M}=\emptyset}$, be the sets of UEs, SCBSs, CUEs and AUEs in the system respectively. Let ${\mathcal{M}(n)}$ and ${\mathcal{M}(l)}$ denote the sets of CUEs in coverage of AUE $n$ and SCBS $l$; ${\mathcal{N}(m)}$ and ${\mathcal{N}(l)}$ the sets of AUEs in coverage of CUE $m$ and SCBS $l$; ${\mathcal{L}(m)}$ and ${\mathcal{L}(n)}$ the sets of SCBSs in coverage of CUE $m$ and AUE $n$. Let ${\delta_{(l,k)}} \in\{0,1\}$, ${\delta_{(l,n,m)}} \in\{0,1\}$ represent association variables for I2D and D2D 
mode selection taking value 1 if the association between SCBS $l$ and UE $k$ or between SCBS $l$, AUE $n$ and CUE $m$, correspondingly is set, and 0 otherwise. Let $\omega_{l_{cue}} \in[0,1]$, $\omega_{l_{aue}}\in[0,1]$ and $\omega_{n_{cue}}\in[0,1]$ accordingly account for MAC scheduling enforced and turn-related penalty weights of SCBS $l$ for served CUEs and AUEs, and of AUE $n$ for attended CUEs.
Achievable rate for CUE $m$ that depends on its mode selection, and for AUE $n$ are captured in \eqref{eq:u_m_long} and \eqref{eq:u_n}; The welfare provided by SCBS $l$ derived from the rates of CUEs and AUEs serviced directly in I2D is captured in \eqref{eq:u_l}.
\begin{subequations}
\begin{alignat}{5}
&r_{cue}(m)=\sum\limits_{\substack{l\in \mathcal{L}(m)}}\Bigg(\omega_{l_{cue}}R_{(l,m)}^{mmW}\delta_{(l,m)}+\nonumber\\
&\left.\sum\limits_{\substack{n\in \mathcal{N}(m)}}\hspace{-2mm}\omega_{n_{cue}}\min\hspace{-0.5mm}\left\lbrace\hspace{-1mm}\frac{R_{(l,n)}^{mmW}}{2},\frac{R_{(n,m)}^{mmW}}{2}\hspace{-1mm}\right\rbrace\hspace{-0.5mm}\delta_{(l,n,m)}\hspace{-1mm}\right)\hspace{-0.8mm},\forall m \hspace{-0.8mm}\in\hspace{-0.8mm} \mathcal{M},\label{eq:u_m_long}\\
&r_{aue}(n)=\sum\limits_{\substack{l\in \mathcal{L}(n)}}\omega_{l_{aue}}R_{(l,n)}^{mmW}\delta_{(l,n)},\:\forall n \in \mathcal{N}, \label{eq:u_n}\\
&r_{scbs}(l)=\sum\limits_{m\in\mathcal{M}(l)}\omega_{l_{cue}}R_{(l,m)}^{mmW}\delta_{(l,m)}+\nonumber\\
&\sum\limits_{\substack{n\in \mathcal{N}(l)}}\omega_{l_{aue}}R_{(l,n)}^{mmW}\delta_{(l,n)},\:\forall l \in \mathcal{L}.\label{eq:u_l}
\end{alignat}
\end{subequations}
The network wide system performance is given by:
\begin{equation}\label{eq:u_system}
\Gamma=\sum\limits_{l\in\mathcal{L}}r_{scbs}(l).
\end{equation}

After collecting control variables $\varphi_i^t$ and $\varphi_i^r$ in vectors $\bm{\varphi}^t$ and $\bm{\varphi}^r$, the  throughput maximization under beamwidth selection for mmWave enabled SCN with relay nodes is given by the following optimization  problem:
\begin{subequations} \label{eq:mainOptprb}
	\begin{align}
	\underset{\bm{\varphi}^t,\bm{\varphi}^r}{\text{Max }}&\Gamma=\sum\limits_{l\in\mathcal{L}}r_{scbs}(l), \label{eq:mainOptprb_a}\\
	\text{s.t.}&\sum\limits_{l\in \mathcal{L}(m)}\hspace{-1.9mm}\delta_{(l,m)}\hspace{-1mm}+\hspace{-3.5mm}\sum\limits_{n\in \hspace{-0.1mm}\mathcal{N}(m)}\sum\limits_{l\in \mathcal{L}(n)}\hspace{-2mm}\delta_{(l,n,m)}\hspace{-0.8mm}=\hspace{-0.8mm}1, \forall m \hspace{-0.7mm}\in\hspace{-0.7mm}\mathcal{M}, \label{eq:mainOptprb_b1}\\
	&\sum\limits_{l\in \mathcal{L}(n)}\delta_{(l,n)}=1, \forall n \in\mathcal{N}, \label{eq:mainOptprb_b2}\\
	&\sum\limits_{m\in \mathcal{M}(l)}\hspace{-2mm}\delta_{(l,m)}+\hspace{-2mm}\sum\limits_{m\in \mathcal{M}(n)}\sum\limits_{n\in \mathcal{N}(l)}\hspace{-1mm}\delta_{(l,n,m)}\geq 1, \forall l \in\mathcal{L}, \label{eq:mainOptprb_c1}\\
	&\sum\limits_{l\in \mathcal{L}(n)}\sum\limits_{m\in \mathcal{M}(n)}\delta_{(l,n,m)}\geq 0, \forall n \in\mathcal{N}, \label{eq:mainOptprb_c2}\\
	&\delta_{(m,l)} \in \{0,1\},\space\delta_{(l,n,m)} \in \{0,1\},\space\forall (l,m,n), \label{eq:mainOptprb_i}\\
	&{\varphi_i^t}{\varphi_i^r}\geq \frac{T_p}{T}{\psi_i^t}{\psi_i^r},\label{eq:mainOptprb_g}\\
	&\varphi_{i_{min}}^t\leq \varphi_i^t\leq\psi_i^t, \varphi_{i_{min}}^r\leq\varphi_i^r\leq\psi_i^r \label{eq:mainOptprb_h}
	\end{align}
\end{subequations}

To tackle the beamwidth selection problem leading to the maximization of welfare, SCN topology and beamwidth bounds related constraints apply. Among the first ones, cons\-traint \eqref{eq:mainOptprb_b1} states that each CUE associates to either an SCBS or an AUE. Constraint \eqref{eq:mainOptprb_b2} indicates that each AUE associates to a single SCBS, and \eqref{eq:mainOptprb_c1} specifies that each SCBS can serve many CUEs/AUEs. While \eqref{eq:mainOptprb_c2} means that AUEs not serving any CUE and AUEs having one or more CUEs served are both allowed provided the rest of constraints are met. Constraint \eqref{eq:mainOptprb_i} explicitly states the binary nature of $\delta$ association variables. Finally, constraints \eqref{eq:mainOptprb_g} and \eqref{eq:mainOptprb_h} relate to the beam width bounds at transmitter $t$ and receiver $r$ of link $i$ as formulated per \eqref{eq:low_beamwidth_product}.

\section{Methodology and Simulation Setup}\label{sec:algorithm+setup}

In order to efficiently solve the problem formulated in Expressions \eqref{eq:mainOptprb_a} through \eqref{eq:mainOptprb_h} a centralized approximative solver based on Swarm Intelligence is used. This paradigm refers to all such methods capable of solving optimization problems via computationally light agents (also called \emph{particles}) that interact with each other by means of simple yet collectively intelligent behavioural policies. Such policies simulate social activity patterns observed in bird flocks or fish schools when they react and adapt to environmental circumstances. The adaptive collective behaviour observed in these species permits them to discover optimum regions within search spaces under a measure of global fitness. This work embraces the computational efficiency of this broad family of algorithms towards solving the addressed beamwidth optimization problem in a centralized fashion.

\begin{figure}[t]
	\includegraphics[width=1\columnwidth]{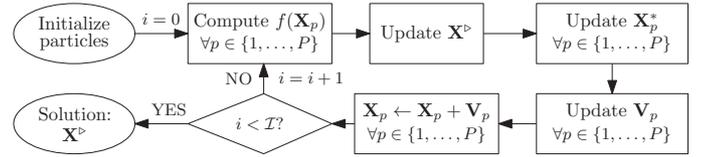}
	\caption{Overall flow diagram of the proposed PSO-based approach.} \label{fig:PSO_flow}
\end{figure}

Many algorithmic approaches relying on Swarm Intelligence have been hitherto proposed in the literature. We will focus on one of the most seminal alternatives, the so-called Particle Swarm Optimization (PSO \cite{Eberhart1995}), which operates by iteratively updating a pool of $P$ candidate solutions $\{\mathbf{X}_p\}_{p=1}^P$ (with $\mathbf{X}_p=(x_1^p,\ldots,x_D^p)$). Following the diagram in Figure \ref{fig:PSO_flow}, the search procedure begins by producing a initial set of particles --which for the problem at hand denote the widths of the transmission and reception beams of the deployed nodes-- and their $D$-dimensional velocity vectors $\mathbf{V}_{p}=(v_1^p,\ldots,v_D^p)$. It should be clear that for the problem at hand, $D=2K-1$ in a system with $K$ nodes, $K-1$ transmission beams (one per link) and $K$ reception beams (one per node). This initialization assigns a fixed value (3\degree) for all beamwidths, whereas each dimensional component of their velocity vectors $\mathbf{V}_{p}$ are drawn uniformly at random from the range [0\degree, 90\degree]. Then the algorithm evaluates the fitness function $f(\mathbf{X})$ given in \eqref{eq:u_system} at each particle location, giving rise to the global best fitness value $f(\mathbf{X}^\triangleright)$, which is achieved by the global best particle location $\mathbf{X}^\triangleright=(x_1^{\triangleright},\ldots,x_D^{\triangleright})$. The best individual position $\mathbf{X}_p^\ast=(x_1^{p,\ast},\ldots,x_D^{p,\ast})$ for particle $\mathbf{X}_p$ is also computed and kept along the algorithm thread. The procedure follows by iteratively refining velocities based on the current velocity, particles' individual best locations, and the best locations of their neighbours as
\begin{equation}
v_d^p \leftarrow \eta v_d^p + \gamma r_{\gamma} (x_d^{p,\ast} - x_d^p) + \xi r_{\xi} (x_d^\triangleright - x_n^p),
\end{equation}
where it should be noted that the second summand in the right-hand expression denotes the difference between the current position and the best position the particle has ever seen, and the third component in the same expression is the difference between the current position and the best position in the whole population of particles. Parameters $\eta$ (inertia), $\gamma$ and $\xi$ permit to control the search behaviour of this heuristic, whereas $r_\gamma$ and $r_\xi$ are realizations of a uniform random variable with support $[0,1]$. The new particle location results from the old one $\mathbf{X}_p$ as $x_d^p \leftarrow x_d^p + v_d^p$, with $d\in\{1,\ldots,D\}$ and $p\in\{1,\ldots,P\}$. This process is repeated until a stopping criterion is met, e.g. a fixed number of iterations $\mathcal{I}$ has been reached. Our implemented PSO approach uses $P=30$ particles, $\eta=0.5$, $\gamma=\xi=1.5$ and $\mathcal{I}=200$ iterations.
\begin{figure*}[t!]
	\centering
	\captionsetup[subfloat]{farskip=2pt,captionskip=1pt}
	\subfloat[N=5  (AUE/CUE)$_{e}$=5/95]{\includegraphics[width=0.66\columnwidth]{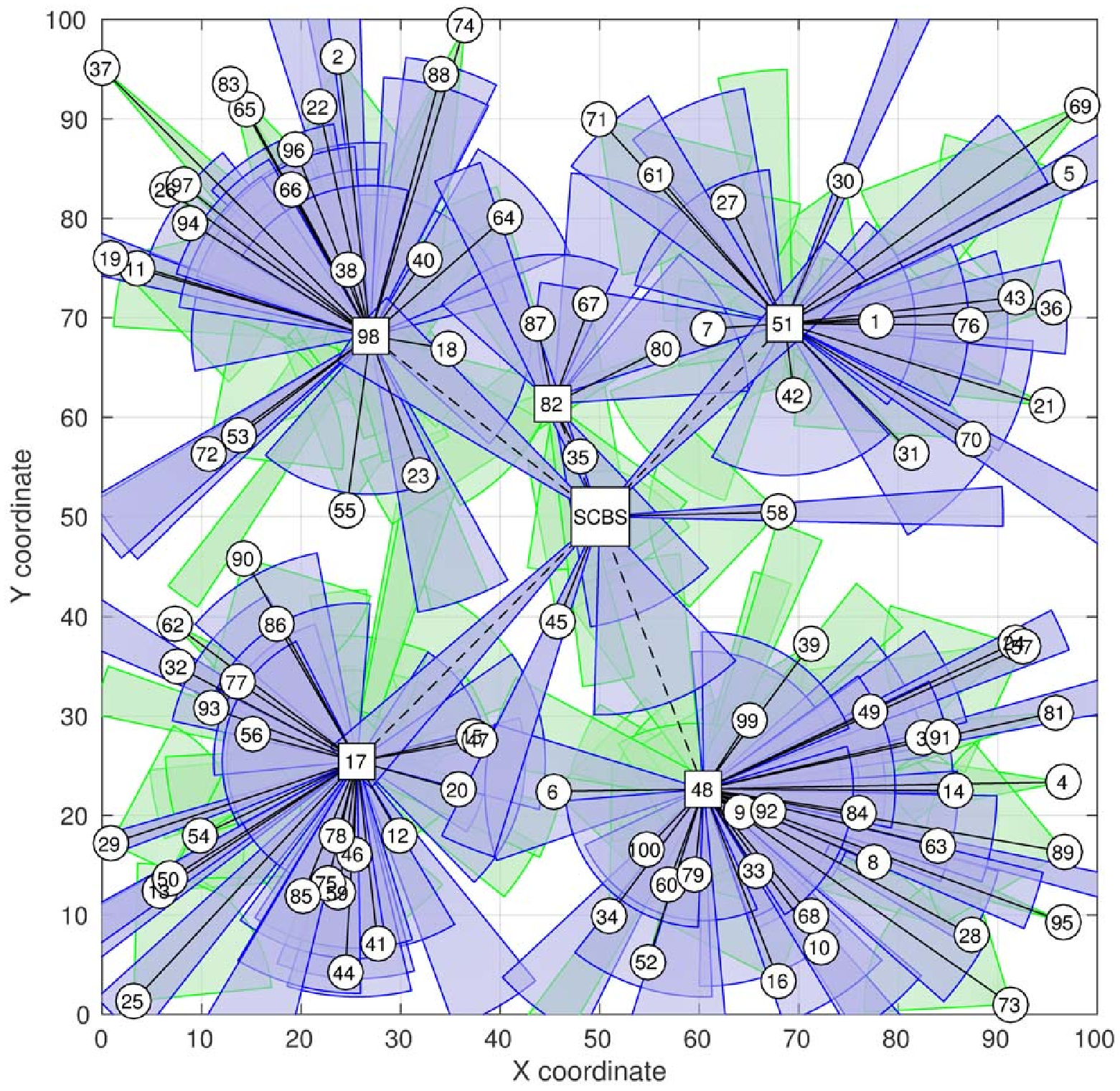}\label{fig:AUE/CUE_5/95}}\hfil
	\subfloat[N=10 (AUE/CUE)$_{e}$=9/91]{\includegraphics[width=0.66\columnwidth]{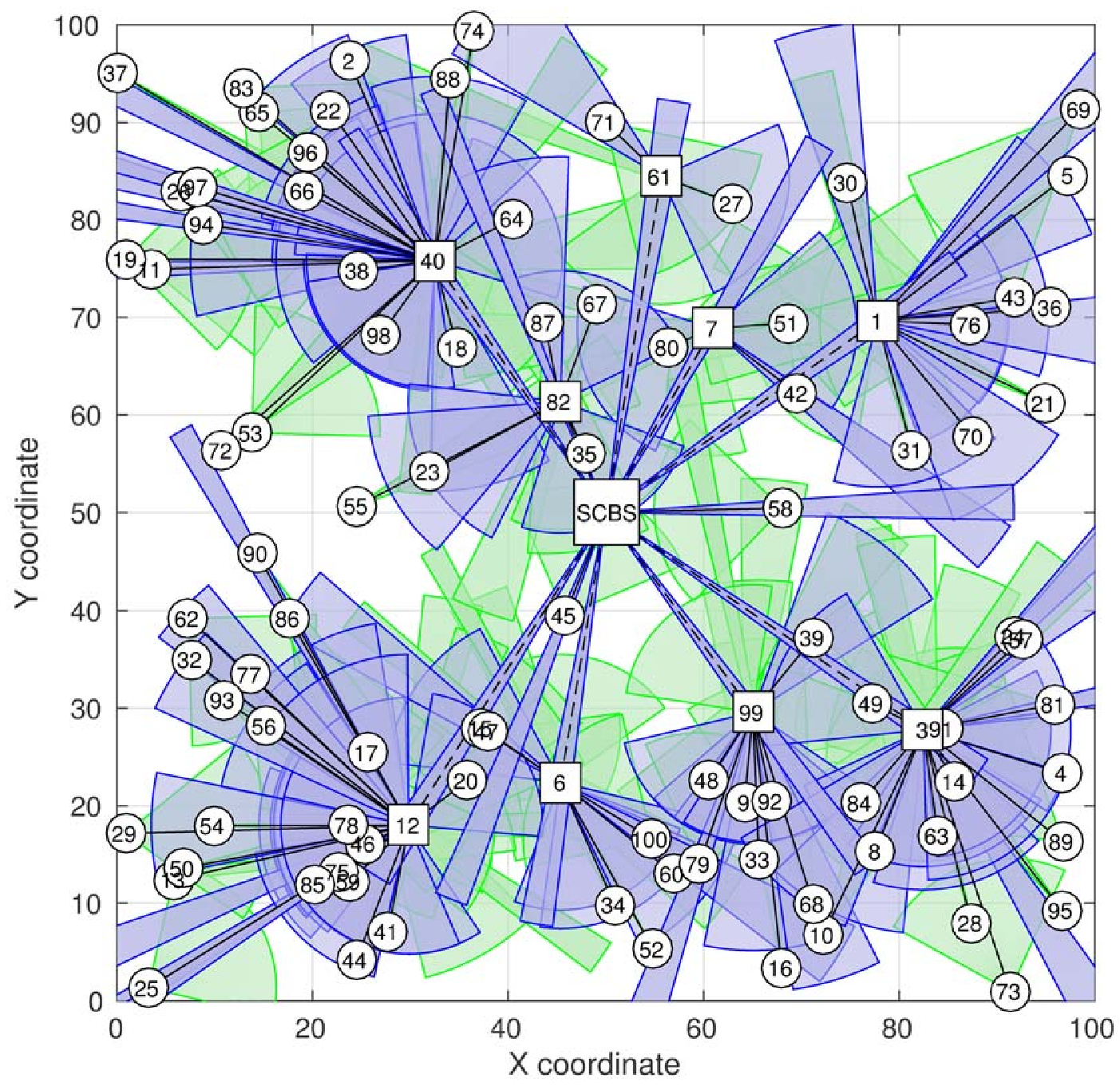}\label{fig:AUE/CUE_9/81}}\hfil
	\subfloat[N=20 (AUE/CUE)$_{e}$=18/82]{\includegraphics[width=0.66\columnwidth]{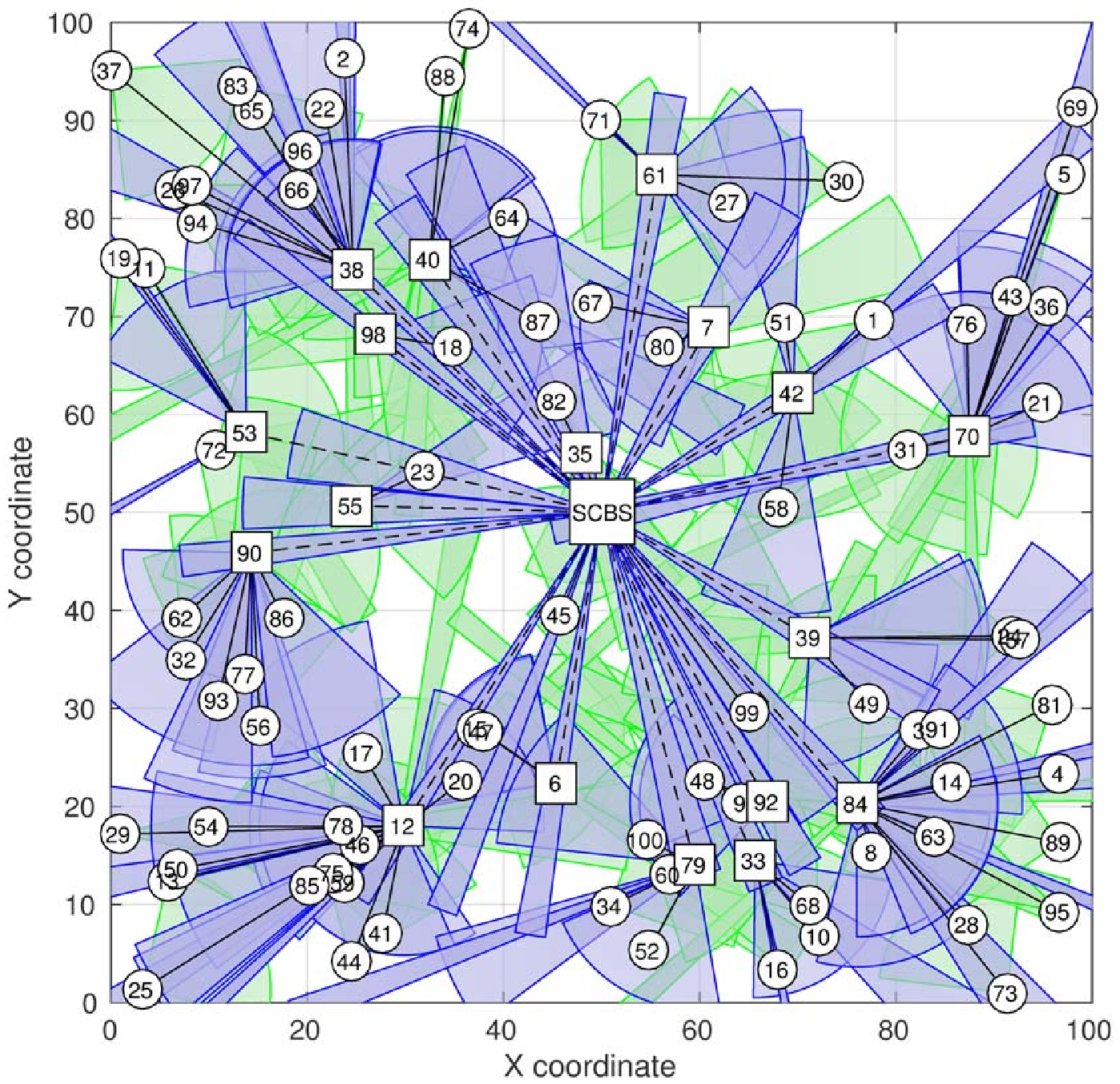}\label{fig:AUE/CUE_18/82}}
	\caption{Resulting optimized beamwidths for nominal AUE/CUE ratios $\in\{5/95,10/90,20/80\}$. SCBS and AUEs are represented as squares, whereas SCBS to AUE links are distinguished from regular links by using dashed lines.}\label{fig:Net_Top_AUE/CUE}
\end{figure*}

\subsection{Simulation Setup}
The performance of the proposed PSO solver when applied over the problem at hand has been assessed for a fixed topology with $K=100$ nodes deployed at random over a square grid of $100\times 100$ meters and $l\triangleq1$ SCBS located in the middle. Different AUE/CUE densities have been emulated over this deployed network by selecting $N$ out of the $K$ nodes to act as relay for CUEs in close proximity. Without loss of generality AUE role assignment and the allocation of CUE nodes to anchors is done under a two-fold criteria: eligible AUE candidates, those in coverage distance for the SCBS, will be named AUE after a search process tags them as those producing best $N$ SINR results --better LOS/interference pro\-perties to communicate with the SCBS--, whereas the remaining CUEs will be connected to the closest among the set of AUEs and the SCBS itself in concordance to current mmWave standards \cite{ieee:802.11ad_part11,ieee:802.15.3.c_part15.3} where criteria of minimum distance association or maximum received signal strength indicator (RSSI) are applied. Correct AUE selection is not trivial since the rate is limited by the backhaul connection from SCBS. Once the association has been performed, nominal AUEs with no served CUEs will revert their status to regular CUEs operating in I2D mode selection. By following this simple procedure a network topology with structural properties meeting the requirements imposed in \eqref{eq:mainOptprb_b1} to \eqref{eq:mainOptprb_c2} can be produced for any nominal value of $n\leq |\mathcal{K}(l)|$, with $\mathcal{K}(l)$ being the UEs in coverage of $l$-th SCBS. In particular simulations for $n\in\{5,10,20\}$ will be hereafter discussed towards quantitatively shedding light on the following points:
\begin{enumerate}
	\item The average rate gains obtained when the transmission and reception beamwidths of the constituent nodes of the network are optimized by means of the proposed PSO approach for different AUE/CUE ratios.
	\item The effect of the number of AUEs in the interference or noise-limited regime that governs the system.
	\item The performance of the proposed PSO-optimized mmWave network subject to TDMA vs. an ideal `all on' MAC.
\end{enumerate}

Simulations parameters for simulated AUE/CUE densities include: a central operation frequency of $f=60$ GHz, $\mathcal{B}_{mmW}=1.2$ GHz, transmit powers of $p_{i}^{SCBS}=30$ dBm and $p_{i}^{UE}=15$ dBm, $N_0=-174$ dBm/Hz, a distance-dependent path-loss model characterized in mmWave by $\mbox{PLE}=2$ \cite{Rappaport2015}, and peak transmit time to slot time $T_p/T = 10^{-2}$. Due to the stochasticity of the PSO search procedure (imposed by $r_\gamma$ and $r_\xi$), performance statistics (represented by median, upper and lower quartiles)
will be computed over $10$ Monte Carlo expe\-riments for each simulated network scenario. MU-MIMO alike operation rendering multiple --in our simulations unlimited-- independent mmWave beams in SCBS $\omega_{l_{cue}}=\omega_{l_{aue}}=1$ will be contemplated. For AUEs, both TDMA leading to $\omega_{n_{cue}}$ inversely proportional to the number of CUEs linked to AUE$_n+1$ and `all on' strategies with $\omega_{n_{cue}}=1$ will be examined.

\section{Results and Discussion}\label{sec:results}

The discussion starts by providing some intuitions on what a priori and a posteriori results are expected to unveil:

\vspace{1mm}- Prior to PSO: a noise-limited regime should prevail in low AUE/CUE ratio scenarios, whereas high AUE/CUE ratios should be more prone to operating under an interference limited regime. And contingent upon named AUEs being in LOS with SCBS and with their served CUEs, adding more AUEs should be beneficial in terms of achieved rates: higher AUE/CUE ratios mean serving fewer CUEs per AUE and thus the effect of a restraining MAC scheduling should be less severe. Nevertheless, it also implies having more simultaneous transmitters, so the effect of interference could counteract rate improvements. The previous intuition could be easily confirmed by comparing results under TDMA in AUEs and under an ideal `all on' MAC for all considered AUE/CUE ratios. Without the burden of the scheduling policy, increasing the number of AUEs --and as a result the feasible sources of interference-- would become advantageous only to provide content from backhaul to CUEs too far from the SCBS.

\vspace{1mm}- Once PSO is in effect: under a fixed AUE/CUE ratio, as the number of iterations of the PSO solver increases beamwidths are refined. Wider beams will be preferred to partially alleviate induced alignment delay and increase throughput, yet that implies 1) increasing potential interference in the main lobe and 2) a reduced received power.

\begin{figure}[t!]
	\captionsetup{farskip=2pt,captionskip=1pt}
	\centering
	{\includegraphics[width=1\columnwidth]{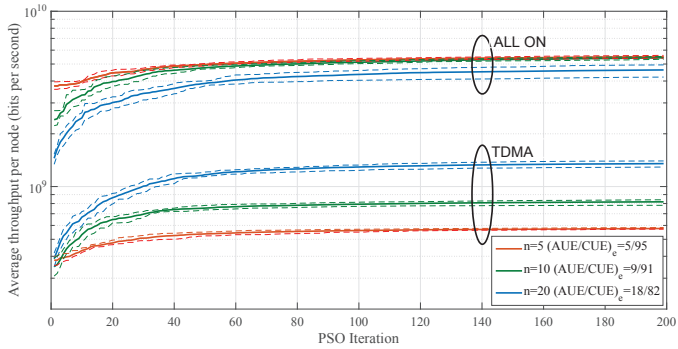}\label{fig:convergence}}\\
	\caption{Evolution of Average Rate under PSO for different AUE/CUE ratios.}
	\label{fig:PSO_Convergence}
\end{figure}

As for simulation results, Fig. \ref{fig:Net_Top_AUE/CUE} graphically represents refined transmission and reception beamwidths for the considered AUE/CUE densities. The evolution of the average throughput per node in Fig. \ref{fig:PSO_Convergence} validates the main hypothesis that Swarm Intelligence is able to efficiently optimize the system welfare. Average interference, SINR and throughput per UE are also provided in Tables \ref{tab:SINR+Interf-SimulValues}-\ref{tab:Throughput-SimulValues} for different AUE/CUE densities and both TDMA and `all on' scheduling.

A detailed look into Tables \ref{tab:SINR+Interf-SimulValues}-\ref{tab:Throughput-SimulValues} evinces that the throughput enhancement comes as a result of sacrificing effective SINR for a fewer alignment delay; and that an under-restrictive scheduling only acts as a multiplicative factor to this extent. A possible explanation is that the alignment's delay contribution to throughput is linear, whereas SINR's is logarithmic. Obtained interference levels for increasing AUE/CUE ratios are consistent with the idea of adding extra interference sources. As the number of transmitters increases how PSO excels at lessening interference is more clear, leading to decrements of up to 66,2$\%$ in TDMA and 97.61$\%$ under `all on' scheduling. However, for low transmitter ratios in TDMA the improvement is very mild or even negative, which suggests that increasing the beamwidths allows for more interferers, even if with smaller gains. The reason being that TDMA scheduling itself imposes a hard limit onto simultaneous transmissions. Noteworthy is to highlight that all interference levels fall within the range of the background noise (-174 dBm/Hz and 1.2 GHz results in around -83 dBm). This is due to the sidelobe gain and the distance between nodes being high enough to impose such interference levels.

It should be remarked here that \cite{Shokri-Ghadikolaei2015d} acknowledges that negligible MUI not necessarily implies noise power being the main bottleneck for throughput when elaborating on the transitional behaviour of mmWave networks. Specifically the effect of other procedures, such as beam-training overhead, noticeably impact achievable performance and, thus, working regime. Therefore, in simulations results, noise, interference, alignment delay or the design of MAC strategy can all affect and ultimately limit the performance of the system. This, indeed, is what can be inferred from the above plots and table: when the impact of the medium access strategy is removed from the throughput calculation (by e.g. resorting to a holistic approach as has been assumed here) the system can be optimized to move from a interference-limited to a noise-limited regime, as the interference levels obtained after PSO clearly show. However, when a practical MAC scheme is implemented, the overall throughput of the system is dominated by interference. All in all, disregarding the MAC approach considered in the system, the reduced SINR level produced by the PSO approach conclude that any interference minimization effort is outgained by a lower alignment delay. It is better --to some extent, as interference is lowered in most scenarios-- to speed up the alignment between nodes even if this comes along with an SINR penalty.
\begin{table}
	\renewcommand{\arraystretch}{1.1}
	\caption{Mean Interference and SINR for effective AUEs}
	\label{tab:SINR+Interf-SimulValues}
	\centering
	\begin{tabular}{|c|c|c|c|c|c|c|}
		\hline \textbf{TDMA}&\multicolumn{3}{c|}{\textbf{Interference (dBm)}}&\multicolumn{3}{c|}{\textbf{SINR (dB)}}\\
		\hline \textbf{AUE$_e$} & Initial & Final &\%& Initial & Final &\%\\
		\hline 5 &-87.54 &-86.93&15,08&69.55 &32.91 &-52,68\\	
		\hline 9 &-84.48 &-84.68&-4,50&69.53 &35.93 &-48,32\\
		\hline 18 &-73.92 &-78.63&-66,19&64.83 &34.68 &-46,51\\
		\hline
		\hline \textbf{`all on'}&\multicolumn{3}{c|}{\textbf{Interference (dBm)}}&\multicolumn{3}{c|}{\textbf{SINR (dB)}}\\
		\hline \textbf{AUE$_e$} & Initial & Final &\%& Initial & Final &\%\\
		\hline 5 &-92.93 &-93.96&-21,11 &72.26 &34.47&-52,30\\	
		\hline 9 &-88.91 &-91.04&-38,76 &71.79 &33.21&-53,74\\
		\hline 18 &-77.74 &-93,96&-97,61&66.87 &36.96&-44,73\\		
		\hline
	\end{tabular}
	\vspace{0.2cm}
	\caption{Mean Throughput per Node for effective AUEs}
	\label{tab:Throughput-SimulValues}
	\begin{tabular}{|c|c|c|c|c|c|c|}
		\hline \textbf{TDMA}&\multicolumn{3}{c|}{\textbf{$\Gamma$/UE (Gbps)}}\\
		\hline \textbf{AUE$_e$} & Initial & Final&\%\\
		\hline 5 &0,3799 &0,5762&51,67\\
		\hline 9 &0,3489 &0,818&134,45\\
		\hline 18 &0,3876 &1,352&248,80\\	
		\hline\hline \textbf{`all on'}&\multicolumn{3}{c|}{\textbf{$\Gamma$/UE (Gbps)}}\\
		\hline \textbf{AUE$_e$} & Initial & Final&\%\\
		\hline 5 &3,748 &5,493&46,56\\	
		\hline 9 &2,406 &5,424&125,44\\
		\hline 18&1,45&4,626&219,03\\	
		\hline
	\end{tabular}

\end{table}

\section{Conclusion and Future Research Lines}

This work has elaborated on the trade-off between alignment delay and interference mitigation and its impact on the overall throughput of relay-based mmWave downlink communications. In particular a centralized beamwidth optimization approach based on Swarm Intelligence has shown that the particular medium access strategy in use is crucial and restricts significantly the overall achievable throughput in mmWave systems. Once the MAC effect is smoothed, the alignment delay has been proven to dominate the effective throughput and cannot be obviated any longer to resort to a traditional SINR centric optimization approach. The latter statement holds whenever an alignment procedure based on exhaustive angular search is implemented. This, in fact, motivates further research aimed at coupling beamwidth optimization with topology formation, as well as deriving novel alignment strategies with increased search efficiency.

\section{Acknowledgements}

This work has been partially funded by the Basque Go\-vernment under ELKARTEK grant ref. KK-2015/0000080.

\bibliographystyle{IEEEtran}

\end{document}